\documentclass[10pt]{article}

\usepackage{amsmath}
\usepackage{amssymb}

\usepackage{graphicx}

\usepackage{cite}

\usepackage{color} 

\usepackage[labelfont=bf,labelsep=period,justification=raggedright]{caption}

\makeatletter
\renewcommand{\@biblabel}[1]{\quad#1.}
\makeatother

\date{\today}

\begin{document}

\begin{flushleft}
{\Large
\textbf{Dynamic spatial and network sampling}
}
\\
Steven K. Thompson$^{1}$, 
\\
\bf{1} 
 Department of Statistics and Actuarial Science, Simon Fraser
 University, Burnaby, BC, Canada
\\
$\ast$ E-mail: thompson@sfu.ca
\end{flushleft}


This paper considers some designs for sampling and interventions in
dynamic networks and spatial temporal settings.  The sample spreads
through the population largely by tracing network links, although
random sampling or spatial designs may be used in addition.  To
investigate the effectiveness of different designs for finding units
on which to make observations and introduce interventions is
investigated through simulations.  For this purpose a dynamic spatial
network model is developed based on simple stochastic processes.  The
sampling processes considered have both acquisition processes and
attrition process by which units are added and removed from the
sample.  The effect of an intervention introduced with a given
sampling design is assessed by the change in the resulting equilibrium
distribution or, in more detail, by the distribution of sample paths
resulting from the intervention strategy.


\section*{Introduction}

Sampling of populations that change in time involves a number of
challenges.  For populations having spatial structure, network
relations, or both, additional challenges arise, along with features
that a sampling design can take advantage of.  Many of the
hard-to-reach human populations are unevenly distributed in geographic
or social space and at least partially hidden from view.  

Dynamic network sampling refers to sampling designs for obtaining
samples in dynamic networks as well as to dynamic designs for
obtaining samples in static networks.  

This paper describes an approach to evaluating the effects of
interventions in epidemics and obtaining effective intervention
strategies using network modeling and a sampling
design approach to introducing potential interventions.  Local
intervention strategies to alleviate the HIV epidemic serve as the
focus.  For the purpose of evaluating combinations of strategies, a
dynamic network model in constructed and simulations with interacting
designs are analyzed.  While relatively simple the modeling approach
is powerful for the purpose.  It includes variable factors of interest
such as clustering, mixing, rates of change in relationships, social
drift and change, stage-specific transmission rates, and changes in
degree distributions.  Interventions are introduced into this
population through link-tracing and combination sampling designs, so
that different ways of introducing them can be compared.  The types of
interventions to be compared individually and in combinations include
ones that exist, such as access to condoms, promotion of reduced
change rates in relationships or in numbers of concurrent
relationships, access to combination antiretoviral treatments and
adherence programs, and seeking earlier testing and treatments, and
distribution programs for potential treatments that are not yet
available, such as vaccines and treatments to clear the virus.  

Flexible sampling designs termed generically dynamic network sampling
are used for introducing the interventions into the population.  The
designs, which use a combination of link-tracing and other designs
such as random sampling are described first in idealized form.  These
designs also serve as models for what happens in the real situations
of contact-tracing, seek and treat, and various respondent-driven
sampling procedures in which the actual procedures used are not
completely controlled.  Finally, but significantly, the class of
designs serves as the model for how a virus such as HIV spreads in the
population.  In the statistical sampling literature a sampling design
is generally characterized by a selection process for bringing units
into the sample.  These designs have in addition an attrition process
by which units are removed from the sample.  The stochastic balance
between the selection and removal processes determine whether the
sample is increasing, decreasing, or fluctuating in an equilibrium
distribution.  With a controlled sampling design this sample size
distribution is determined by the investigators.  In a design such as
seek and treat the largest possible sample size might be desired, but
reaches natural limits based on increasing difficulties of contact
tracing or participants dropping out of a program.  In the case of a
virus network sampling design, we can seek to intervene with our own
design to decrease the rate of spread and drive the equilibrium
distribution to the lowest possible level.

Static network models have been extended to the dynamic situation
fairly recently by a number of approaches.
\cite{snijders1996stochastic,snijders2001statistical,snijders2002markov,snijders2005models}
and \cite{snijders2007modeling}
developed an interesting class of network evolution models based on
behavioral characteristics of actors/nodes such as tendencies toward
reciprocity, transitivity, homophily, and assortative matching.  At
the same time he developed inference methods to estimate model
parameters from incomplete longitudinal data.  A summary of this
work together with a review of other approaches to dynamic network
modeling is contained in \cite{snijders2010introduction}.
\cite{krivitsky2010separable} present a dynamic network model based on
exponential random graph models.  A recent summary of statistical
network models is provided by \cite{goldenberg2009survey}.

The social space approach to network modeling is introduced in 
\cite{hoff2002}, using latent
space models for (static) networks,  and a simple dynamic extension of this
approach is described by \cite{Sarkar:2005:DSN:1117454.1117459}.
Spatial point processes with clustering tendencies have been used as
the basis for simulations of stochastic graph models (static rather
than dynamic) with which to evaluate sampling designs and inference
methods by
\cite{thompson2006aws,targetedwalks2006,thompson2011adaptive}.

Design-based inference methods for link-tracing designs in networks
have been developed in
\cite{thompson2006aws,targetedwalks2006,thompson2011adaptive}.  The
current approach to likelihood-based inference with link-tracing
designs was introduced in \cite{thompson2000model}, developed for
Bayes inference in \cite{chow2003estimation}, with Markov chain Monte
Carlo computational Bayes inference used for adaptive web sampling
designs in Kwanisai, M. (2005) [Estimation in Link-Tracing Designs with Subsampling.
Ph.D. thesis. The Pennsylvania State University, University Park, PA,
USA] and 
\cite{kwanisai2006estimation}.  This approach was further
developed used with additional models in
\cite{handcock2010modeling}. Issues of inference in network epidemic
models are additionally discussed in \cite{welch2011statistical}.

\cite{salehi2010empirical} describe empirical likelihood based
confidence intervals for adaptive cluster sampling.
\cite{deville2006indirect} analyze weighting systems for estimation
in indirect sampling, a class of adaptive network designs.

Approaches combining design and model based methods include the random
walk asymptotics based estimators used with respondent driven sampling
\cite{heckathorn1997respondent,heckathorn2002respondent},
\cite{salganik2004sampling}, \cite{volz2008probability}, and
\cite{targetedwalks2006}.  A different
approach is used in \cite{félix2004combining} and
\cite{félixmedina2006combining}.  A different approach still
combining design and model based methods is developed in
\cite{gile2011network}.

An approach bring network effects to epidemic compartment models by
having different groups with different network degrees is described in
\cite{house2011insights}.  Properties of networks in which
relationships shift preferentially that are missed by static network
epidemic models are examined in \cite{fefferman2007disease}.  The
stability of casual contact and close contact patterns over time was
studied \cite{read2008dynamic} using diary based methods with 49
volunteers, finding that the close contacts tended to be more stable
than casual contacts.  Epidemic threshold properties of simple dynamic
network models in which degree says constant but neighbors exchange,
that is, identity of partners change instantaneously at random times
are described in \cite{volz2009epidemic}.  Compartmental models are
modified to add some network effects in \cite{bansal2007individual}.
Network structure and change patterns in relation to individual
behaviors were investigated in \cite{potterat1999network} for
syphilis among young people and HIV among drug users, finding in both
studies that spread of disease was associated with network cohesion,
in the form of separation of components or local density of connections.  \cite{gupta1989networks} examine network effects in
compartmental models by including a contact or mixing matrix into the
model, comparing assortative mixing patterns, in which individual's
contacts tend to be within their own group and dissortative patterns,
in which contacts tend to be between groups.  Early work in modeling
the dynamics of the HIV epidemic includes 
\cite{may1987transmission, may1988transmission} and 
\cite{anderson1986preliminary}.   The importance of concurrent
relationships in the spread of the HIV epidemic is investigated  in
\cite{watts1992influence} and \cite{morris1997concurrent}.

The models for dynamic spatial network populations and interacting
designs developed in work also have some relationship to the
literature on evolutionary dynamics.  \cite{nowak2006} provides a summary
of models of species interactions, epidemics, and selection based on
systems of partial differential equations.  \cite{champagnat2006unifying}
describe recent work on stochastic point process models built on top
of that approach. Among the many cases of ecological systems
exhibiting the sort of spatial, temporal, and network patchiness
addressed by the models and designs of this paper, a good example
is described by \cite{rudicell2010impact}.

\cite{phillips2013increased} (Supplement 1) modeled the individual variability of
adherence over time for an individual as well as the variability
between individuals.  Their model, used for the purpose of estimating
parameters, is individual based but not network based except to the
extent of having assumed degree distributions.  
\cite{kamp2010} describes a modeling approach based on a set of
partial differential equations with the addition of some temporal network
aspects such as degree distributions in which contacts change,
describing mean behavior for infinite network size and approximate
behavior for moderate size.  \cite{cohenet2012} presents
opposing views of different authors on the relative importance of
early stage transmission.  
\cite{rocha2011simulated}. \cite{rocha2012epidemics} describe a
network of 50,185 sexual contacts between 6,642 escorts and 10,106 sex 
buyers as reported on a Web discussion forum. Dynamic epidemic models
showing selective advantage of virulent strains in early stage of an
epidemic and advantage shifting to less virulent strains, which allow
their host to survive longer, as the epidemic matures were compared to
laboratory studies with colonies two competing strains of bacteria
\cite{Berngruber2013}, finding agreement with the model predictions.

\section*{Methods}

\subsection*{Dynamic network model}

To investigate the effectiveness of different sampling and
intervention strategies, a model producing a dynamic population
network is required.  For this purpose, the model does not necessarily
need to include every possible detail of a population, but should
include any relevant factor that affects the effectiveness of a design
or strategy.  We model the population as fluctuating yet stable, that
is, existing in a stochastically fluctuating equilibrium, so that it
provides a base for trying different intervention strategies and
following the results through time, comparing those outcomes with
those without the intervention.  We want to model to be rich enough so
that we can investigate the effects of different amounts of
clustering, rates of change, mixing between groups, degree
distributions, and other relevant parameters.  The model is
constructed in layers, each layer a simple stochastic process.  First
there is a time changing spatial process.  This is a spatial point
cluster process, in which the clusters or social groups drift in a social space and
nodes within groups change position relative to each other
over time.  This approach is in the spirit of the the social space
models of \cite{hoff2002} and \cite{Sarkar:2005:DSN:1117454.1117459}but with the
additional attributes such as clustering, mixing, social change, sex,
degree dynamics, and
population dynamics involving birth and death processes needed for the
sampling strategy investigations.  The
motion of the points is based on small increments of diffusion yet is
stable in the sense of maintaining the clustering over time.  Links
form and dissolve probabilistically over time based on distance
between nodes and other factors including sex of node, current degree
of node.  The stochastic duration of nodes, or conversely the rate of
change in relationships, is governed by a renewal process.  In
addition, an insertion and deletion process accounts for births,
deaths, immigration, and emigrations of nodes from the study region
and includes a population dynamics to keep the population size and
even group sizes in an equilibrium distribution.  This is the base
population to which the interacting sampling strategies are applied.

\subsection*{Sampling designs}

The type sampling designs we consider for obtaining an ongoing sample
of units from the ever-changing population includes both a selection
process and an attrition process.  The selecting process is dominated
by link-tracing, but to a lesser extent, and also at the initiation of
sampling, units may be selected at random or by some other standard
design such as a spatially distributed one.  At time $t$ in the
sampling process we have a sample $s_t$ of units or nodes.  We trace a
set of links out from that sample and add the linked nodes to the
sample.  Additional nodes may be selected by simple random sampling or
some type of time-location sampling, though the probability of this
may be small relative to the link-tracing selection of new units.

A simple but versatile class of designs in this type uses conditional
Bernoulli link tracing.  In these designs a link $(i,j)$ from node $i$
to node $j$ is traced with probability $p_{ij}$, independently, given
current values at that time step, of the tracing of other links.  The
tracing probabilities can depend on values of the origin and
destination node or of the relationship between them and can change
over time.  The number of links followed at any time step is random
and at many times may be zero.  In addition to depending on individual
nodes and links, the probabilities of tracing can depend on
characteristics of the entire sample including current sample size.

Other designs in the class include tracing at most one link out from
the sample at a given time step, or selecting a random sample without
replacement of a target number of the nodes, tracing all of the links
out if there are fewer of them than the target.

The attrition process by which nodes are removed from the sample
includes, unavoidably in the dynamic setting, any deletions of nodes
from the population due to deaths or emigrations.  In addition, the
attrition procedure may include removal of nodes from the sample
without their deletion from the population.

Active sample units are those from which links, if available, can be
traced.  The sample may also include inactive units, such as people
who are retained in the sample for treatment or follow-up interviews
but from whom links will no longer be traced.  The sample may be
with-replacement or without-replacement.  The concepts active and
with-replacement are generalized to continuous variables between 0 and
1, where a value less than 1.0 represents dampening of the probability
of tracing a link from that node or of selecting it again after its
removal from the sample.  

Sample size and its increase or decrease is controlled by the balance
between the selection and attrition processes.  In a seek and treat
design it may be desired to have sample size as large as possible and
to retain members in treatment.  In a monitoring design for
surveillance it may be desired to maintain a sample size near a target
value, even as the individuals making up the sample change over time.
This can be accomplished by adjusting the selection probability
downward if the sample size is above target and adjusting it upward if
sample size is below target.  Similarly, rates of removal from the
sample can be adjusted based on current sample size.  With the seek
and treat designs sample sizes tend to come into a stochastic
equilibrium distribution over time as it becomes harder to find new
cases or trace from them, as participants drop out of the sample, or
budget constraints impose limits.  

To understand what controls the equilibrium distribution of sample
size over time it is useful to view some aspects of the network
geometry of a link-tracing sample in a dynamic network.  At any time
point the sample has a node volume, which is the number of nodes in
it, and a link surface, which is the number of links from nodes in the
sample to nodes outside the sample.  As sample size increases the
ratio of surface to volume tends to decrease, with most designs of the
type we consider.  Partly this is because as sample size increases in
a finite population and links are followed at some point there will be
fewer links to follow.  But the decrease in ratio typically occurs
much earlier.  One reason is that link-tracing design of the types
considered here generally select units having high degree (many links) with
high probability.  As those units and their contacts come into the
sample many of the links become contained in the sample and there
become over time fewer links pointing out of the sample.  Second, a
node with degree $d$ coming in to the sample through tracing of a link
from a sample node, will have up to $d - 1 $ of its links pointing out
of the sample.  Units selected into the sample at a given time tend to
have higher degree than is average for units in the sample and higher
degree out of the sample than the average for sample units at that
time.  The more time it spends in the sample, the smaller the
proportion becomes of its links that point out of the sample.  For the
sample as a whole, as the sampling progresses and sample size becomes
larger, a higher proportion of links from sample nodes are internal to
the sample, leaving a smaller proportion on the surface.  Clustering
of links in the population also affects the pattern of surface in
relation to volume during the course the sampling.  The speed of
growth of the sample depends on the number of links on its surface and
the tracing rates of these surface links.  The rate of attrition from
the sample, on the other hand, is related to the volume of the sample
and the deletion and removal rates associated with those units.  Thus
as surface to volume ratio increases, the rate of acquisition relative
to attrition decreases and the sample size comes into a stochastic
equilibrium distribution.  These effects are striking in simulation of
various types of sampling designs, both human-designed ones and
natural ones.

With link-tracing designs, the probability that a node is selected
into the sample is usually higher for nodes having more links, or more
or more links pointing in to them in the cased of directed links.
While there are exceptions to this, such as the uniform walk designs
in Thompson (2006b), in almost all cases higher selection
probabilities for nodes with higher degrees result naturally from the
way the sample is selected.

With a dynamic population such as modeled here, in which nodes drift
over time in an underlying social space, links form in relation to
social distance and change over time, and nodes and links have some
clustering tendencies, the speed of sampling in relation to this
change has an effect on the type of sample that results.  If the rate
of link-tracing is very fast, then selection of the next $k$ nodes,
for $k$ some fixed number, might tend to be all or most in the same
cluster or highly connected.  If the rate of link-tracing is very
slow, the next $k$ nodes are more likely to be spread out in social
space and less highly connected with each other and include selections
from different clusters.

\section*{Example - HIV Epidemic}


A local population or community through which an epidemic of HIV
spreads conceived as a dynamic network, with nodes representing people
and links representing the types of relationships between people
through which the virus can transmit.  We think of the virus as having
a dynamic network sampling design that selects a sample of people by
following links, so that the current sample of the virus is the set of
individuals in the population who are infected.  By infecting a person
the virus makes an intervention that affects not only the state of the
person's health but affects the temporal network structure of the
population by affecting survival rates and with them, network
connections over time.  Per-act transmission rates, while generally
low, are variable depending on stage of infection, specific nature of
the contact act, and other factors.  Since there is at present no
treatment to consistently clear the virus from a person's body the
attrition process of the sampling design occurs with mortality or,
from the perspective of the local modeling, when an individual moves
out of the study community.  In addition to link-tracing, nodes can
come into the virus' sample through immigration of an infected person
into the study region or through ``random'' events such as a resident
individual traveling outside the region and returning with an
infection.

In terms of the sample design parameters used by the virus in
spreading the epidemic, the design is without-replacement if we are
considering only the increase in sample size as the measure of
prevalence of the disease in the population.  On the other hand,
multiple infections do occur and can be modeled as with-replacement
selection, which may of practical importance in terms of
recombination events between different strains of virus.  The
transmission rate at reinfection may be lower than at an initial
infection.  Highest transmission rate occurs with early stage of
infection, having a variable duration of several weeks.  This is
followed by a lower rate for a long period during chronic stage and
possibly a higher rate later.  In terms of design parameters this
corresponds to the most active units in the sample, in terms of
tracing rate, being those recently selected, with ``active'' being a
continuous variable that changes over time.  Membership in the sample
and stage affect deletion rate through mortality and possibly affect
emigration out of a study population also.  Further, an increase in
tracing rate is associated with an increase in deletion rate, when
comparing different strains and different stages.
Tracing/transmission rate from one node to another is affected by
values of the origin node, the destination node, and the link or
relationship between them.  Finally, attrition from the sample occurs
only with deletion from the population, so long as there is no
practical cure for HIV available.  With development and distribution
of a functional cure, however, direct removal of a person from the
sample would become a possibility.

Interventions to counter the spread of the virus are similarly
conceived as being distributed to individuals in the community with a
dynamic network sampling design that uses random or time-location
sampling or a combination of procedures.  Widely used procedures for
bringing interventions, such as seek and treat designs using contact
tracing and respondent driven sampling methods using coupons can be
evaluated in this way.  The ideal is to bring interventions in such a
way that they benefit not only the sample individuals to which they
are applied but the wider community by decreasing further spread.  A
seek and treat design might for example start with cases that are
tested.  If the test is positive, the links from that node are with
some probability traced.  When a link is traced, the node to which it
leads is if possible tested.  If the test is positive, interventions
including combination antiretroviral drugs to improve health and
decrease the rate of transmission onward can be prescribed along with
counseling.  If the test result is negative, interventions could
include counseling and provision of barrier methods to decrease the
probability of transmission in.  When a vaccine becomes available that
is even partially effective this is where it could be offered.

In terms of the sample geometry, tracing rates, and temporal speed of
sampling described in the previous section for dynamic network
sampling designs, the virus HIV appears to have a dual strategy.  At
times it can spread very fast, appearing in the simulations as a local
explosion.  And it can persist for long periods with low rates of
transmission, while the sample over time spreads out socially and
relationships change, until it encounters favorable conditions and
ignites a new local explosion.  One factor that makes this pattern
possible is the high rate of transmissibility in the brief early stage
of infection compared to the much longer lasting but lower
transmission rate of the chronic stage.  In an expansion in a cluster
of high average degree, when a new node is infected, it enters the
first stage.  With the high transmission rate and locally high number
of links around there is the possibility of one or more additional
transmissions while the newly incident node is still in early stage.
The result is that some, though not all, of the transmissions during
the period of rapid expansion are from nodes in early stage, and the
higher transmission rate of early stage catalyzes the local
expansion. If early stage transmission rate is no higher than the
chronic stage rate, local expansions still occur but require a higher
density of links in a local cluster to have an effect.  A local
expansion cluster can also be catalyzed by a high latent transmission
rate over its links.  This could happen for example in a local
concentration of a contributing infection such as a genital ulcer
disease.

The most prevalent contributing factor to local explosive expansion
is perhaps somewhat surprising and results from the dynamic network
sampling design characteristics.  At incidence, the person newly
infected with HIV has on average a higher degree than is average for
people in the population.  This is a design property of link-tracing
designs of this type.  This is especially pronounced in early stages
of the epidemic but persists even at equilibrium levels.  More
notably, in terms of surface and volume properties of the sample, the
newly incident person tends to have a higher proportion of his or her
links pointing to partners outside of the sample, that is to
uninfected partners, compared to the average degree out for all
infected cases.  The degree-out at incidence tends to be higher even
compared to the degree-out that the same individual will have after
being in the sample for some time.  This is because over time more of
the persons contacts come to be infected, either from that individual
or another since those contacts are in a high-risk area of 
social/network space.   With the high degree-out of an incident node
coinciding with the high transmission rate of early stage, the effects
of the two factors multiply to increase the rate of spread of the
virus and contribute to the pattern of local explosions.  Once a local
explosion has run its course there tends to be a period of low surface
to volume ratio, and what links there are on the surface tend to
be mainly from nodes in chronic stage, having low transmission rate
but relatively long survival time.  Those are the conditions under
which a dynamic network sample tends to spread out the subsequent new
selections in social/network space.

\section*{Acknowledgments}
This work was supported by the National Science and
  Engineering Research Council

\bibliography{refs2013}



\end{document}